# Pressure induced superconductivity on the border of magnetic order in MnP

J.-G. Cheng[1,2†*], K. Matsubayashi[2†], W. Wu[1†], J. P. Sun[1], F. K. Lin[1], J. L. Luo[1], and Y. Uwatoko[2]

[1] Beijing National Laboratory for Condensed Matter Physics and Institute of Physics, Chinese Academy of Sciences, Beijing 100190, China

[2] Institute for Solid State Physics, University of Tokyo, 5-1-5 Kashiwanoha, Kashiwa, Chiba 277-8581, Japan

**Abstract**

We report the discovery of superconductivity on the border of long-range magnetic order in the itinerant-electron helimagnet MnP via the application of high pressure. Superconductivity with $T_{sc} \approx 1$ K emerges and exists merely near the critical pressure $P_c \approx 8$ GPa, where the long-range magnetic order just vanishes. The present finding makes MnP the first Mn-based superconductor. The close proximity of superconductivity to a magnetic instability suggests an unconventional pairing mechanism. Moreover, the detailed analysis of the normal-state transport properties evidenced non-Fermi-liquid behavior and the dramatic enhancement of the quasi-particle effective mass near $P_c$ associated with the magnetic quantum fluctuations.

Extensive investigations over last decade have uncovered the quantum criticality as a universal phenomenon connecting with many difficult problems in modern physics.[1, 2] For example, the most distinguished problem of unconventional superconductivity as found in several distinct superconducting systems including the heavy-fermion, organic, cuprates, and the iron-based superconductors can be generally described in the framework of antiferromagnetic quantum critical point (QCP).[3-6] The close proximity of superconductivity to a magnetic instability suggests that the critical spin fluctuations would play a crucial role for mediating the Cooper pairs.[5, 7] On the other hand, to realize a magnetic QCP should provide an effective approach for searching new classes of unconventional superconductors. This is well illustrated by the recent discovery of pressure-induced superconductivity in CrAs,[8, 9] the first Cr-based unconventional superconductor.[10] This discovery has left manganese (Mn) the only 3d element that does not show superconductivity among any Mn-based compounds; the strong magnetism of Mn is commonly believed to be antagonistic to superconductivity. Therefore, it is highly interesting to explore whether superconductivity can emerge near a magnetic QCP in the Mn-based compounds.

For this purpose, the itinerant-electron helimagnet, MnP,[11] with a much reduced moment of ~1.3 $\mu_B$ per Mn has attracted our attention as a good starting point to approach a magnetic instability. At ambient condition, MnP adopts an orthorhombic *B31*-type structure with lattice constants $a$ = 5.26 Å, $b$ = 3.17 Å, and $c$ = 5.92 Å, respectively. Here we use the *Pnma* setting $c > a > b$ through this paper. In the absence of a magnetic field at ambient pressure, MnP undergoes two successive magnetic transitions upon cooling:[11] a transition from paramagnetic to ferromagnetic (FM) state at $T_C$ = 291 K, and then a second transition to a double helical state at $T_s \approx$ 50 K. In the FM state between $T_C$ and $T_s$, the Mn spins are aligned parallel to the orthorhombic *b* axis, and the ordered moment is about 1.3 $\mu_B$ per Mn atom. In the double helical state at $T < T_s$, the Mn spins rotate in the *ab* plane with the propagation vector **q** along the *c* axis.[12] Earlier hydrostatic pressure studies[13-15] on MnP have revealed negative pressure coefficient for both $T_s$ and $T_C$. In addition, the high-pressure ac magnetic susceptibility measurements by Banus[15] indicated that the FM transition changes to an antiferromagnetic (AFM) type at $P \geq$ 3 GPa. However, the pressure and temperature ranges in the previous studies were far from the magnetic QCP. The recent developments of high-pressure techniques[16-19] that can maintain a good hydrostatic pressure conditions over 10 GPa allow us to reinvestigate this problem and finally lead to astonishing findings. We found that MnP becomes superconducting below $T_{sc} \approx 1$ K when its long-range magnetic order is just suppressed by the application of high pressure ~8 GPa. This discovery makes MnP the first manganese-based superconductor, and the close proximity of superconductivity to a magnetic instability suggests an unconventional pairing mechanism.

Needle-shaped MnP single crystals used in the present study were grown out of a Sn flux. Measurements of resistivity and ac magnetic susceptibility under hydrostatic pressures up to 10 GPa were performed by using various high-pressure techniques. Without specification, the resistivity in the present study was measured with the current applied along the growth direction, which is found to be the orthorhombic *b* axis, the easy-magnetization direction. Details about the crystal growth and high-pressure measurements can be found in the Supplemental Materials (SM).

The resistivity $\rho(T)$ data given in Fig. 1 illustrate an overall evolution with pressure of the above-mentioned two magnetic transitions. Fig. 1(a) shows the *b*-axis $\rho(T)$ data under various pressures up to 10.7 GPa measured with a palm cubic anvil cell. In agreement with the previous reports[20], $\rho(T)$ at zero pressure display a kink anomaly at the FM transition, $T_C = 291$ K, which can be defined clearly as a sharp peak in the $d\rho/dT$ curve, Fig. 1(b). The $\rho(T)$ and $d\rho/dT$ curves at $P = 2.8$ GPa keep essentially similar features as those at ambient pressure, except that $T_C$ has been shifted down to ~250 K. Upon further increasing $P = 5.0$ GPa, however, the temperature profile of $\rho(T)$ and $d\rho/dT$ exhibit distinct features. The $\rho(T)$ curve at 5.0 GPa was separated into two segments by a clear inflection point at ~200 K, above which resistivity increases linearly with temperature; correspondingly, the $d\rho/dT$ curve at 5.0 GPa in Fig. 1(b) displays a step-like anomaly at the inflection point followed by a broad maximum centered around 110 K. As noticed by Banus[15] and discussed below, it is most probable that the FM transition changes to an AFM type for $P > 3$ GPa. For this reason, we label the transition temperature as $T_m$ for $P > 3$ GPa hereafter. The $\rho(T)$ and $d\rho/dT$ curves at 6.4 GPa are similar to that at 5.0 GPa, except that the magnetic transition broadens up and moves down to ~ 120 K. Upon further increasing $P = 7.4$ GPa, nevertheless, $\rho(T)$ changes again with the concave curvature restored for $T > T_m = 70$ K, which is manifested as a relatively broad peak in the d$\rho$/d$T$ curve in Fig. 1(b). Above this pressure, no clear anomaly can be discerned any more in the $\rho(T)$ curves, which are typical for correlated metals showing the resistivity saturation behavior. In this pressure range, the d$\rho$/d$T$ curves shows only a broad hump corresponding to the gradual curvature crossover with temperature in the $\rho(T)$ curve.

Fig. 1(c) shows the $c$-axis resistivity, $\rho_c(T)$, measured in a piston-cylinder cell. As reported earlier, $\rho_c(T)$ at ambient pressure exhibits a clear dip anomaly at $T_s$ because the helical magnetic structure propagates along the $c$ axis. As shown clearly in Fig. 1(c), $T_s$ decreases monotonically with pressure and vanishes completely at ~ 0.9 GPa.

From the above resistivity data, we can see that the application of high pressure quickly eliminates the double helical state, and first reduces the FM transition at $T_C$, and then, most likely, changes it to an AFM type for $P$ > 3 GPa. The magnetic transition monitored by the anomaly in $\rho(T)$ eventually vanishes completely around $P_c \approx$ 8 GPa, where the most striking change take place at low temperatures. As shown in Fig. 2(a), we start to see a resistivity drop below 1 K at 7.6 GPa, and a more pronounced resistivity drop with an onset temperature of ~1 K is clearly observed near the critical pressure $P_c \approx$ 7.8 GPa, which signals the possible occurrence of superconductivity. With further increasing pressure, this anomaly shifts to lower temperatures. Although zero resistivity can be hardly reached even when the applied electrical current is reduced to 10 μA, the ac magnetic susceptibility, 4π$\chi$(T), shown in Fig. 2(b) provides strong evidence for the occurrence of superconductivity near $P_c$. In perfect agreement with the $\rho(T)$ data, the diamagnetic signal appears below $T_{sc} \approx$ 1 K at 7.6 GPa, and the superconducting shielding fraction reaches ~95% of the sample volume at 7.8 GPa. A further increment of pressure to 8.6 GPa lowers $T_{sc}$ to below 0.5 K, while the superconducting shielding fraction keeps nearly constant. Such a perfect diamagnetic response rules out the possibility of filamentary superconductivity or impurity phases. However, the absence of zero resistivity could be caused by the imperfect sample quality or the pressure inhomogeneity. Especially, the nonhydrostatic pressure conditions would have a more profound impact on the electrical transport properties. The observation that the superconductivity disappears quickly after the magnetic transition vanishes completely highlights that the pressure-induced superconductivity has an intimate correlation with the magnetic critical point.

To gain further insights into the superconducting state, we obtained the upper critical field $\mu_0H_{c2}$ from the field dependence of $\rho(T)$ and $4\pi\chi$ at $P = 7.8$ GPa, Fig. 3(a, b), and plotted $\mu_0H_{c2}$ as a function of $T_{sc}$ in Fig. 3(c). Here, we define $T_{sc}$ as the temperatures corresponding to 50% resistivity drop and 1% ac susceptibility drop. As shown in Fig. 3(c), $\mu_0H_{c2}$ versus $T_{sc}$ is better described by a linear fitting, which yields a $\mu_0H_{c2}(0) = 0.33(1)$ T, and an initial slope of -$\mu_0\mathrm{d}H_{c2}/\mathrm{d}T_{sc}|_{Tsc} = 0.34(1)$ T/K. The obtained $\mu_0H_{c2}(0)$ allows us to estimate the Ginzburg-Landau coherence length $\xi = 315$ Å according to the relationship: $\mu_0H_{c2}(0) = \Phi_0/2\pi\xi^2$.

The transition temperatures, $T_C$, $T_m$, $T_s$, and $T_{sc}$, obtained from the above resistivity measurements are mapped into the temperature-pressure phase diagram shown in Fig. 4(a). As can be seen clearly, the application of high pressure reduces continuously the magnetic transition temperatures, $T_C$ and then $T_m$, and eventually suppresses the magnetic order around $P_c \approx 8$ GPa. Superconductivity with a maximum $T_{sc} \approx 1$ K emerges and exists within a narrow pressure range near the critical pressure $P_c$. Such a superconducting T-P phase diagram is remarkably similar with that of heavy-fermion superconductors, such as CeIn$_3$ and CePd$_2$Si$_2$,[3] in which a magnetically mediated mechanism is believed to play a dominant role for forming Cooper pairs. In the case of MnP, the situation becomes more complicated in that the application of pressure may alter the nature of the FM transition to an AFM type around 3 GPa.

To follow directly the evolution with pressure of the magnetic transitions at $T_C$ and $T_s$, we also measured the ac magnetic susceptibility, $\chi'(T)$, under pressure. As shown in Fig. S2, they are manifest as a sudden jump and drop, respectively, at ambient pressure, and the ferromagnetic state corresponds to the in-between plateau. In agreement with the resistivity data shown in Fig. 1, $T_C$ decreases continuously, and $T_s$ vanishes completely around 1.4 GPa. Surprisingly, a new two-step transition emerges above 1.4 GPa and increases quickly with pressure. This new transition denoted as $T^*$ is also added in Fig. 4(a). As can be seen, $T^*$ increases quickly and seems to merge with $T_C$ around 3 GPa. Since no anomaly in $\chi'(T)$ can be discerned at $P = 4.2$ GPa, the magnetic transitions reflected in resistivity above 3 GPa should correspond to an AFM one. However, whether this AFM state is similar with the low-pressure double helical phase below $T_s$ deserves further studies, especially with the help of the high-pressure neutron diffraction techniques. But, we provide here further evidences that are in favor of an antiferromagnetic QCP at $P_c$, and thus an unconventional pairing mechanism for the observed superconductivity in MnP.

Fig. 5(a) shows the low-temperature normal-state resistivity in the form of $\rho$ versus $T^2$. As can be seen, the Fermi-liquid behavior is followed nicely for the $\rho(T)$ data below 3 GPa. However, $\rho$ versus $T^2$ curves in the pressure ranges of $4 < P < 7$ GPa and $P > 7$ GPa exhibit a super- and sub-linear behavior, respectively, which will give rise to a resistivity exponent $n > 2$ and $n < 2$, respectively, in the $\rho(T) = \rho_0 + BT^n$ fitting. The pressure dependence of exponent $n$ obtained from the power-law fitting is shown in Fig. 4(b). As can be seen, $n$ first increases from 2 for P < 3 GPa to ~3 at 4 GPa, and then decreases progressively with pressure, until it reaches about 1.5 at the critical pressure $P_c$, above which it starts to increase again. The jump of exponent $n$ around 3 GPa might reflect the pressure-induced FM to AFM transition. The exponent $n = 1.5$ near $P_c$ is consistent with the theoretical predictions for the incoherent scattering of quasiparticle near a 3D antiferromagnetic QCP.[3] Although the above fitting evidenced a dramatic enhancement of the resistivity coefficient B around $P_c$, the variation of exponent $n$ prevents a quantitative comparison. Then, we performed a linear fit to the $\rho$ versus $T^2$ curves, i.e., $\rho = \rho_0 + AT^2$, in the low-temperature limit as shown by solid lines in Fig. 5(a). As shown in Fig. 5(b, c), the residual resistivity $\rho_0$ displays a sharp peak just before $P_c$ and the A coefficient displays a pronounced peak centered at $P_c$. Since the A coefficient is proportional to the effective mass of charge carriers via $A \propto (m^*/m_0)^2$, the significant enhancement of A, signals a dramatic enhancement of effective mass associated with the suppression of magnetic order. The observations of the $\rho \propto T^{1.5}$ behavior and the dramatic enhancement of A coefficient near $P_c$ have been regarded as characteristic signatures of an antiferromagnetic QCP in strongly correlated metallic systems, and provide important clues for the unconventional nature of the pressure-induced superconductivity in MnP.

Finally, it is interesting to mention another itinerant-electron helimagnet MnSi, which has been extensively studied in the context of a ferromagnetic QCP.[21] Although anomalous metallic properties have been observed near the QCP, no superconductivity has been observed down to very low temperatures. This comparison highlights the importance of the antiferromagnetic spin fluctuations associated with a QCP for the observed superconductivity in MnP. The mechanism of pressure-induced ferromagnetic-to-antiferromagnetic transition in MnP thus deserves further studies. Goodenough has argued that such a pressure-induced ferromagnetic-to-antiferromagnetic transition is due to bandwidth broadening in the narrow d-band system.[15, 22]

In summary, we have found that the itinerant helimagnet MnP becomes superconducting below $T_{sc} \approx 1$ K when its long-range magnetic order is completely suppressed by the application of high pressure around the critical pressure $P_c \approx 8$ GPa. The close proximity of superconductivity to a magnetic instability suggests an unconventional pairing mechanism. The present finding of the first Mn-based superconductor breaks the general wisdom about the Mn's antagonism to superconductivity. We hope that this discovery will stimulate more work on searching for Mn- and other transition-metal-based superconductors with a higher transition temperature.

## Acknowledgements

We thank Y. P. Wang, J. Q. Yan, J. S. Zhou, N. Mori, F. Steglich, Q. Si, P. J. Sun, P. Dai, and P. Coleman for enlightening discussions. Work at IOP/CAS was supported by the National Science Foundation of China (Grant Nos. 11304371, 11025422), the National Basic Research Program of China (Grant Nos. 2014CB921500, 2011CB921700), and the Strategic Priority Research Program (B) of the Chinese Academy of Sciences (Grant Nos. XDB07020100, XDB01020300).

Work at ISSP/UT was partially supported by Grant-in-Aid for Scientific Research, KAKENHI (Grant Nos. 23340101, 252460135), and the JSPS fellowship for foreign researchers (Grant No. 12F02023).

[†]These authors contributed equally to this work.

*E-mail: jgcheng@iphy.ac.cn

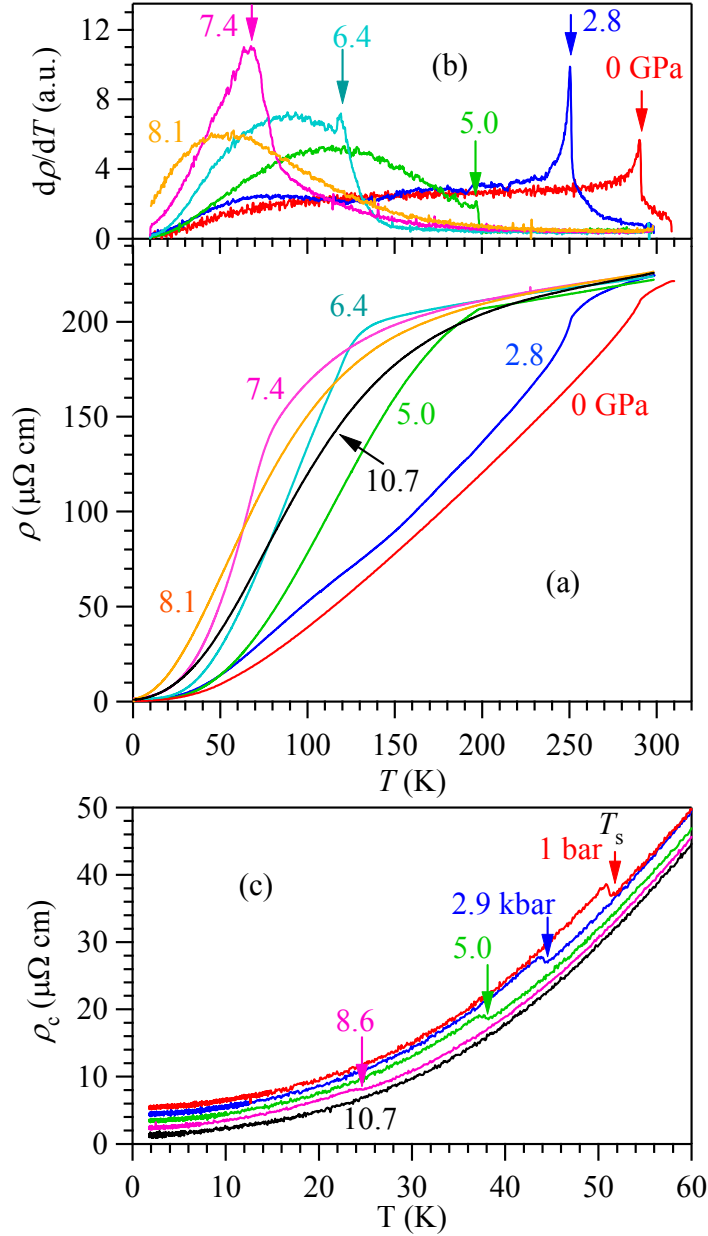

**Fig. 1** (Color online) (a) Resistivity $\rho(T)$ and (b) the temperature derivative $d\rho/dT$ of the MnP single crystal under various hydrostatic pressures up to $P = 10.7$ GPa highlighting the variation with pressure of the magnetic transition indicated by the arrows. (c) The $c$-axis $\rho(T)$ data at low temperatures highlighting the evolution with pressure of the double helical transition at $T_s$.

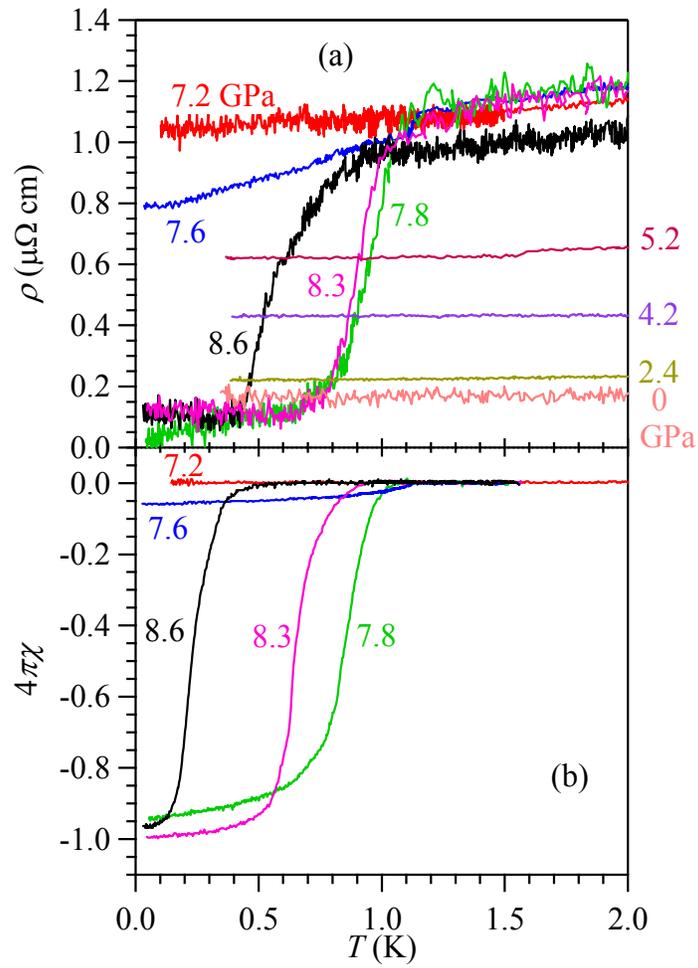

**Fig. 2**(Color online) Temperature dependence of (a) the resistivity $\rho(T)$ and (b) ac magnetic susceptibility $4\pi\chi$ on MnP under various pressures near the critical pressure.

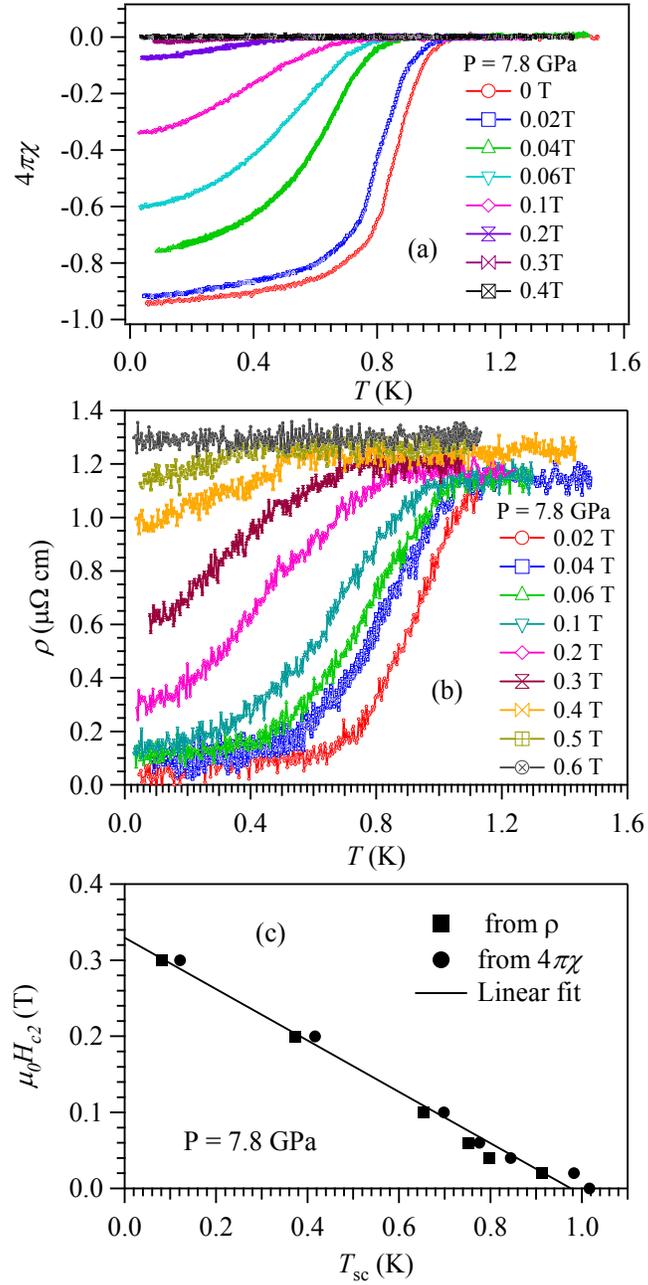

**Fig. 3** (Color online) Temperature dependence of (a) ac magnetic susceptibility $4\pi\chi$ and (b) resistivity $\rho(T)$ under different magnetic fields at $P = 7.8$ GPa. (c) Temperature dependence of the upper critical field $\mu_0H_{c2}$ for MnP at 7.8 GPa. The solid line in (c) is a linear fitting, which gives $\mu_0H_{c2}(0) = 0.33(1)$ T, and a slope -0.34(1) T/K.

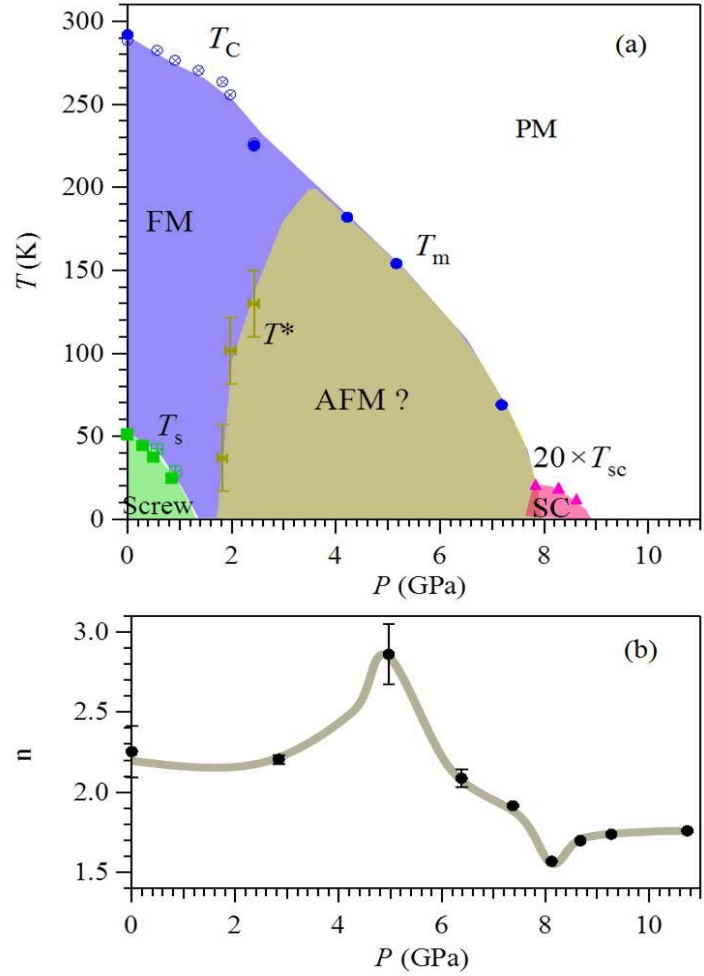

**Fig. 4** (Color online) (a) Pressure dependences of the magnetic transition temperatures, $T_C$, $T_m$, $T^*$, $T_s$, and the superconducting (SC) transition temperature $T_{sc}$ for; $T_{sc}$ has been scaled by a factor of 20 for clarity. (b) Pressure dependence of the exponent n obtained from the power-law fitting $\rho \propto T^n$ to the low-temperature resistivity data.

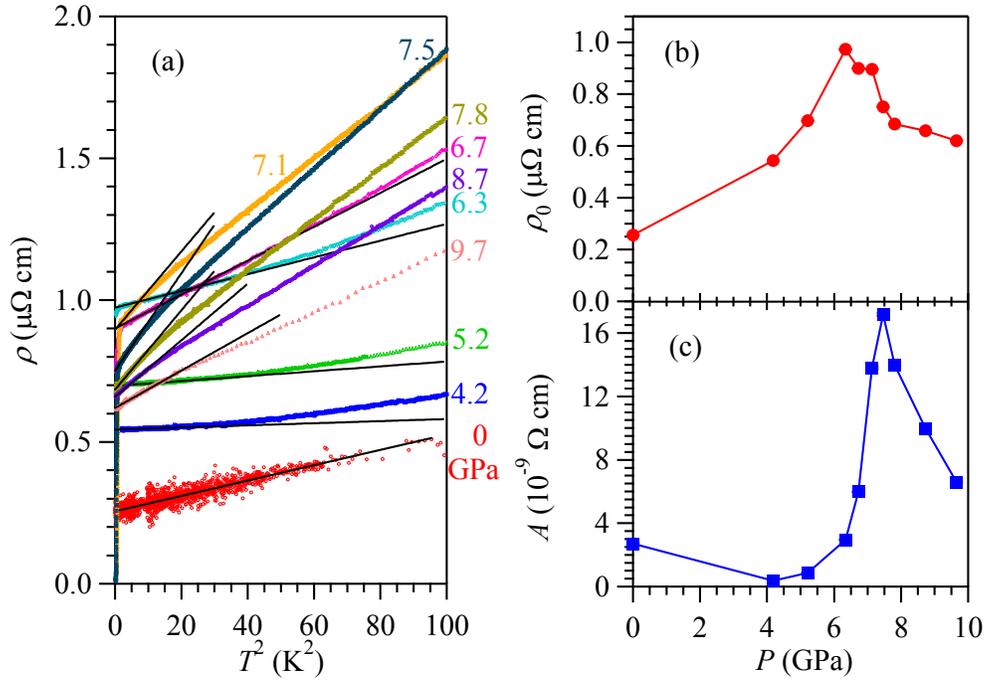

**Fig. 5** (Color online) (a) The normal-state resistivity in the plot of $\rho$ versus $T^2$ for MnP under various pressures up to 9.7 GPa. The solid lines represent the linear fitting in the low-temperature limit. (b,c) The obtained residual resistivity $\rho_0$ and the A coefficient as a function of pressure.

# Supplemental Materials

1.  **Sample preparation and characterizations** MnP single crystals used in the present study were grown out of a Sn flux. The starting materials with a molar ratio of Mn: P: Sn = 1 : 1: 10 were put in an alumina crucible, which was then sealed in a quartz ampoule under a reduced pressure of $10^{-4}$ torr. The quartz ampoule was first heated up to 650°C for 10h, held there for 8h, and then heated up to 1150°C, held for 6h, and then slowly cooled down to 600°C in a period of 200 h. At this temperature, the liquid Sn flux was filtered by centrifuging before the ampoule was cooled to room temperature. The obtained MnP single crystals have a needle shape with typical dimensions 0.2× 0.2 × 3~5 mm$^3$. Powder X-ray diffraction at room temperature confirms that the obtained crystals are single phase with an orthorhombic crystal structure. Before the physical-property measurements, the crystals were further washed with diluted HCl to remove any remaining Sn flux.

2.  **High-pressure techniques** A total of five samples with residual resistivity ratio RRR≡ $\rho(300K)/\rho(2K)$ ranging from 800 to 1100 have been studied under hydrostatic pressures over 10 GPa. Resistivity of two samples (labeled as #1 and #2) was measured with the Palm cubic anvil cell (PCAC) in two separate runs. Another two samples (#3 for resistivity and #4 for ac magnetic susceptibility) were measured simultaneously in an opposite-anvil-type cell (OAC). All the above resistivity measurements were carried out with the current applied along the growth direction, which, in most cases, is confirmed by Laue X-ray diffraction to be the orthorhombic *b* axis, the easy-magnetization direction. The *c*-axis resistivity on the fifth (#5) sample was measured with a conventional piston-cylinder cell (PCC) up to 2.2 GPa.

    The PCAC consists of a cluster of six anvils converging onto the center gasket, in which the sample is immerged in the liquid pressure transmitting medium (PTM) contained in a Teflon capsule. The three-axis pressurization geometry together with the adoption of liquid PTM in PCAC ensures relatively good hydrostatic pressure conditions over 16 GPa. The PCAC can be cooled down to ~0.5 K with a continuous-flow $^3$He cryostat. The pressure values used in the present study for PCAC were based on the low-temperature pressure calibration by monitoring the superconducting transition of Pb. Details about the PCAC and the pressure calibrations can be found elsewhere. [1-3]

Designed by K. Kotigawa *et al*,[4] the opposite-anvil-type cell consists of a pair of opposed tungsten-carbide anvils pressing on a gasket made of a nonmagnetic NiCrAl alloy. In comparison with the traditional opposite anvils such as the modified Bridgman cell, this cell has a substantially improved sample space (~7 mm$^3$) that allows simultaneous measurements of multiple samples. In addition, the compact size of the pressure cell allows easy attachment to a dilution refrigerator so as to reach much lower temperatures necessary for our present study. Moreover, we sealed the Ar gas as the PTM to obtain good hydrostaticity. The pressure was determined from the superconducting transition of Pb.

The PCC is similar to the commonly used one,[5] and can be inserted into the Quantum Design PPMS. It was made from nonmagnetic BeCu and NiCrAl alloys. Glycerol was used as the PTM. The pressure was determined from the superconducting transition of Pb.

The resistivity was measured by the conventional four-probe method. An LR-700 ac resistivity bridge was used to perform the measurement. An electrical current $I = 1$ mA was employed for measurements above 10 K, while the current was decreased down to 10 μA for low-temperature measurements, especially in the dilution refrigerator.

Ac magnetic susceptibility was measured with a mutual induction method at a fixed frequency of 317 Hz with a modulation field of about 1 Oe. The diamagnetic signal due to the superconductivity transition was estimated by comparing to the diamagnetic signal of Pb, which served as a pressure manometer, with nearly the same size as the MnP sample.

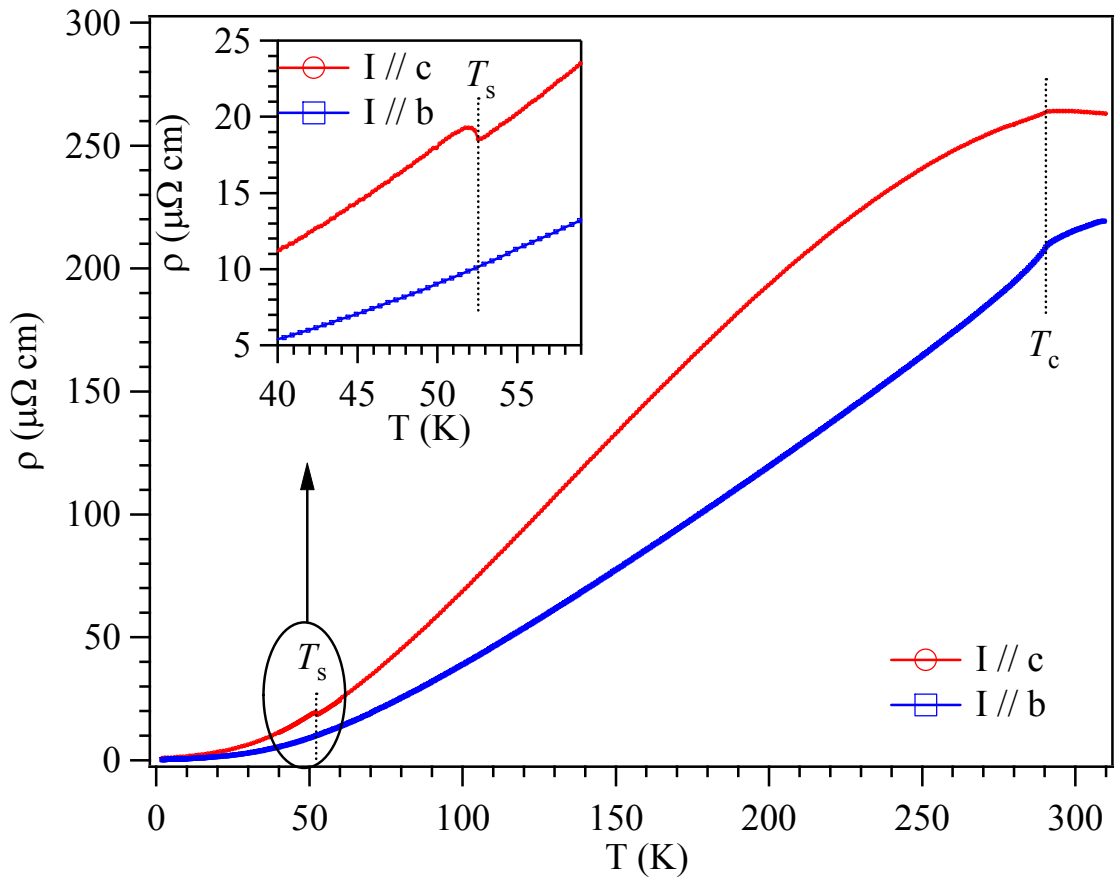

**Fig. S1** Anisotropic resistivity of MnP single crystal with the electrical current applied along the orthorhombic b and c axis, respectively. Because the helical order propagates along the c axis, the double helical transition is only visible along the c-axis resistivity; the transition temperature $T_s$ = 52.5 K is among the highest one ever reported, suggesting the high-quality of the single crystals.

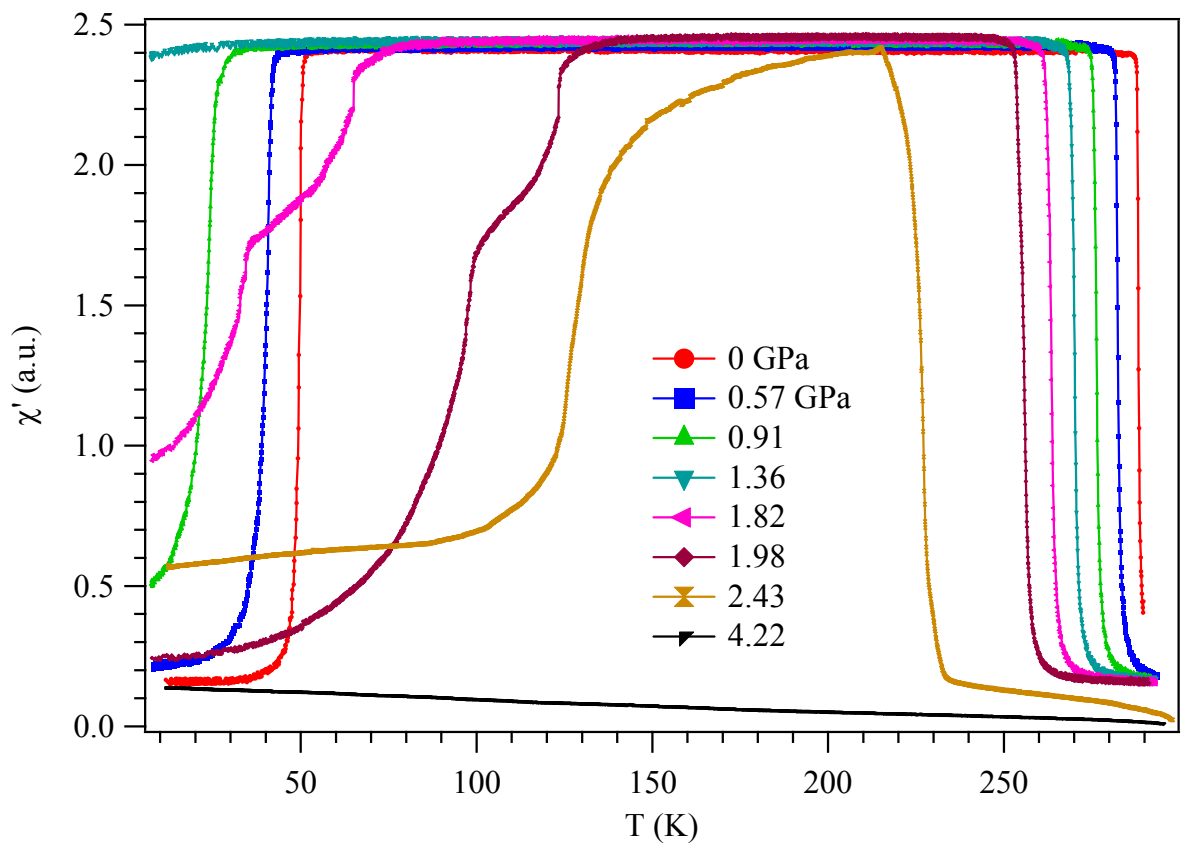

**Fig. S2** Temperature dependence of the ac magnetic susceptibility measured under various pressures.